\documentclass[10pt,a4paper,twoside]{article}
\usepackage{baltlat6}
\usepackage{epsfig}

\begin{document}
\ \ \vspace{-0.5mm}

\setcounter{page}{111}

\vspace{-2mm}

\titlehead{Baltic Astronomy, vol.\,19, 111--119, 2010}

\titleb{DISK GALAXY MODELS DRIVEN BY STOCHASTIC\\ SELF-PROPAGATING STAR
FORMATION}

\begin{authorl} \authorb{T.~Mineikis}{1,2} and
\authorb{V.~Vansevi\v{c}ius}{1,2} \end{authorl}

\begin{addressl} \addressb{1}{Vilnius University Observatory,
\v{C}iurlionio 29, Vilnius LT-03100, Lithuania; e-mail:
tadas.mineikis@ff.stud.vu.lt}
\addressb{2}{Center for Physical Sciences and Technology, Institute of Physics,
Savanori\c{u} 231, Vilnius LT-02300, Lithuania}
\end{addressl}

\submitb{Received 2010 June 2; accepted 2010 July 1}

\begin{summary} We present a model of chemical and spectrophotometric
evolution of disk galaxies based on a stochastic self-propagating star
formation scenario.  The model incorporates galaxy formation through
the process of accretion, chemical and photometric evolution treatment,
based on simple stellar populations (SSP), and parameterized gas
dynamics inside the model. The model reproduces observational data of
a late-type spiral galaxy M\,33 reasonably well.
Promising test results prove the applicability of the model and the
adequate accuracy for the interpretation of disk galaxy properties.
\end{summary}

\begin{keywords} galaxies:  evolution -- galaxies:  disk, individual
(M\,33) \end{keywords}

\resthead{Disk galaxy SSPSF models}{T.~Mineikis, V.~Vansevi\v{c}ius}

\sectionb{1}{INTRODUCTION}

Over the recent decades galaxy evolution models became sophisticated and
based on large state of art codes (Hensler 2009).  Despite this, the
main processes controlling the evolution of galaxies, e.g., star
formation rate and star formation feedback must be taken into account
more carefully.  Therefore, there are still advantages of using relatively
simple models of galactic evolution based on parameterized physical
processes.  Due to the fast computation of these models, one can explore a
wide range of galactic parameters and galaxy properties.

A part of such simple models apply a stochastic self-propagating star
formation (SSPSF) scenario, which has been used with success to explain
the flocculent spiral patterns in late-type disk galaxies (Gerola \&
Seiden 1978).  Initially these models were applied to explain galactic
disk spiral patterns by applying the percolation phenomena to star
formation propagation.  More advanced models include gaseous disks and
perform galaxy modeling self-consistently.  Successful attempts have
been made to explain properties of dwarf galaxies:  Gerola et al.
(1980) have explained a highly diverse star formation rate in these
systems as a characteristic behavior of SSPSF models on a spatially
small disk.  SSPSF models were developed further by introducing
anisotropic propagating star formation probabilities, to account for
different galaxy morphologies (Jungwiert \& Palou{\v s} 1994), and
physical groundings for the self-propagating star formation process
(Palou{\v s} et al. 1994).  By incorporating SSPSF ideas into the models
of disk galaxy evolution, Sleath \& Alexander (1995) were able to
reproduce the Kennicutt-Schmidt law (Kennicutt 1998).  Later on,
following a rapid evolution of computational power, galaxy evolution
models have been developed mainly by applying N-body methods.

We have developed a model based on the ideas of Seiden \& Gerola (1982)
with updated input physics and modern-day knowledge of the star
formation process in disk galaxies.  This model was applied to interpret
observational data of the M\,33 galaxy.

\sectionb{2}{THE MODEL}

We use the disk galaxy model scheme proposed by Seiden \& Gerola (1982)
and supplement it with a prescription of physical processes
characterizing star formation and galaxy evolution.  The disk of a
galaxy is subdivided into regions (cells) of characteristic size, which,
to some degree, are independent entities evolving according to chemical
evolution scenarios.  The most practical representation of such a disk
structure is a two dimensional (2-D) polar grid (Figure~1), which
makes the implementation of the galaxy rotation curve simple and
produces realistic galaxy disk models.

\vskip5mm

%%%%%%%%%%%%%%%%%%%%%%%%%%  FIGURE 1

\vbox{\centerline{\psfig{figure=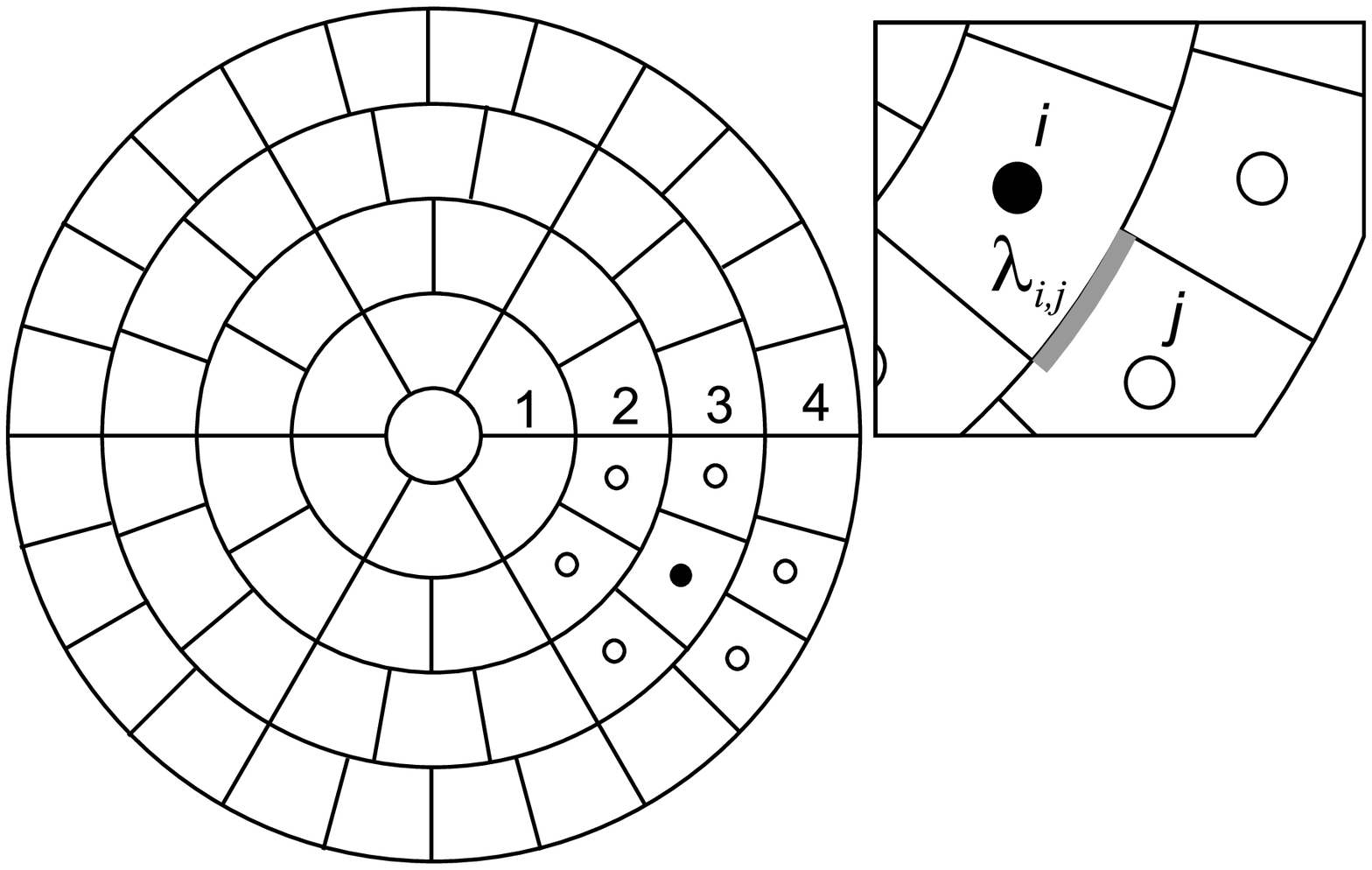,width=70truemm,angle=0,clip=}}
\vspace{0.5mm}
\captionb{1}{The galaxy disk model subdivided into cells. The
neighboring cells (open circles)
 to a cell under consideration (filled circle) and the fraction of its
perimeter ($\lambda_{i,j}$) contacting with a particular
neighbor are marked.}}

\subsectionb{2.1}{Model geometry}

A galaxy is approximated by a 2-D disk subdivided into concentric rings of
the same width (see Figure~1).  A ring with a running number $i$
contains $6\times i$ cells.  This division produces cells of the same
area and perimeter, except for the central cell with an area
smaller by a factor of 3/4.  Cells are the basic structure elements in
the model.  The main parameter describing galactic disks is the number of
rings.  Each ring rotates according to a given rotation curve, which
remains constant during the simulation procedure.

\subsectionb{2.2}{Disk formation}

The simulated galaxy consists of two main parts, i.e., the galaxy itself and
the reservoir, where all the gas is initially located.  The
galaxy formation proceeds gradually by accreting gas from the reservoir.
The rate of accretion onto the galaxy disk, $A(t,R)$, gradually decreases in time and is assumed to be proportional to gas density in the reservoir at a particular galactocentric distance, $R$:
\begin{equation}
A(t,R)=A(0,R)\cdot\exp\left(-\frac{t}{\tau_{\rm acc}}\right),
\end{equation}
where $\tau_{\rm acc}$ is an accretion timescale and $A(0,R)$ is an
initial accretion rate at the galactocentric distance $R$.

\subsectionb{2.3}{Star formation prescription}

A cell is able to experience star formation events (SF events) of two
types: spontaneous and stimulated.  The spontaneous star formation
process is assumed to be related to collisions of giant molecular clouds
(GMCs).  However, we do not consider the formation of GMCs explicitly,
and assume that the number of GMC collisions in the cell is proportional
to $\sigma_{{\rm gas},i}^2$.  Therefore, the likelihood that a cell
experiences a spontaneous SF event is proportional to $(\sigma_{{\rm
gas},i}/\sigma_{\rm SP})^{2}$, where $\sigma_{\rm SP}$ is a parameter
controlling spontaneous star formation, assumed to be equal to
$320\,M_{\odot}\,\textrm{pc}^{-2}$, i.e., double the density of GMCs in
M\,33 (Bolatto et al. 2008).

Stimulated star formation occurs when the cell $i$, which forms stars in the
present time step, induces SF events in the neighboring cells during the
next time step.  This process can be parameterized by the likelihood of 
stimulated star formation, $L_{\rm ST}$, defined as the average number
of new cells in which SF events can be induced by the SF event in the $i$
cell.  Therefore, the likelihood of stimulated star formation in the
neighboring cell $j$ is proportional to $L_{\rm
ST}\,\times\,\lambda_{i,j}$, where $\lambda_{i,j}$ is the fraction of the
$i$ cell's perimeter contacting the cell $j$ (see Figure~1).  In our
model we assume $L_{\rm ST}=2$ (Seiden \& Gerola 1982).

Galaxy disks gradually grow due to gas accretion, therefore, initially
within the entire galaxy and later on at the disk edge star formation
is weak due to the critical gas surface density, $\sigma_{\rm C}$, which is
assumed for star formation. This density reduces the likelihood
of a SF event by a factor of $\sigma_{{\rm gas},i} / \sigma_{\rm C}$,
where $\sigma_{{\rm gas},i}$ is the average gas surface density of a
particular cell.

Schaye (2004) models support the critical surface gas density,
$\sigma_{\rm C} = (3-10)\,M_{\odot}$ pc$^{-2}$, independent of the
galactocentric distance.  We applied $\sigma_{\rm C} = 7\,M_{\odot}$
pc$^{-2}$, a value adopted for the investigation of the disk evolution
in our Galaxy (Chiappini et al. 2001).

During a SF event in a cell lasting one time step (e.g., 10\,Myr), the
fraction of gas converted to stars is defined by star formation
efficiency,
$\epsilon$:
\begin{equation}
\epsilon = \epsilon_0 \cdot \left( \frac{\sigma_{{\rm gas},i}}
{\sigma_{{\rm gas},0}} \right),
\end{equation}
where $\epsilon_0$ is the efficiency of star formation at a gas surface
density used for calibration, $\sigma_{{\rm gas},0}$.
 A linear dependency on
$\sigma_{{\rm gas},i}$ for self-regulated star formation was suggested
by K\"oppen et al.  (1995).  In our model we assume $\sigma_{{\rm
gas},0} = 10\,M_{\odot}\,\textrm{pc}^{-2}$ derived for the
Milky Way galaxy (Wolfire et al. 2003), which produces the observed star
formation efficiency in GMCs (Myers et al. 1986), $\epsilon_0\sim0.02$.

In the next time step after a SF event in the cell, star formation
ceases due to the energy injected by high-mass stars.  Another SF event
in the same cell becomes highly improbable since some time (the
refractory time) is needed to settle and cool the disturbed hot gas. 
Throughout the simulation we use a constant refractory time of 100~Myr, a
value derived for irregular galaxies (Quillen \& Bland-Hawthorn 2008).

\subsectionb{2.4}{Chemical evolution}

The cells in the galaxy model experience discrete star formation bursts.
During the burst in a particular cell, gas (considered to be well-mixed
within the cell) is converted into stars, and the formed stellar
population
can be represented satisfactorily by a simple stellar population (SSP)
approach.
Using SSP properties, calculated with the P\'{E}GASE software (v.~2.0,
Fioc \& Rocca-Volmerange 1997; see Table~1 for the parameters), and
following the track of a cell's star formation history we compute
the chemical
evolution of the cell $i$. The equation of chemical evolution includes the
three main contributions:
\begin{equation}
\frac{\Delta(Z_i\cdot \sigma_{{\rm gas},i})}{\Delta t}=A(t,R)\cdot Z_0+
\sum_{j}F_{i,j}\cdot Z_{i,j}+\sum_{k}G_{i,k}\cdot Z_{i,k}.
\end{equation}
Here the left-hand side represents the change in the metal content of
the cell $i$.  The first right-hand side term is a contribution by the
accreted primordial gas with metallicity $Z_0$ from the reservoir at the
galactocentric distance $R$.  The second term is a sum of gas flows
($F_{i,j}$) from all neighboring cells to the cell $i$ (see below).  The
last term is the sum of metal contributions through all SSPs formed in
the cell $i$ ($G_{i,k}$ is the expelled gas from stars in the stellar
population $k$ with metallicity $Z_{i,k}$).  The evolution of gas
content is described by the equation:
\begin{equation}
\frac{\Delta \sigma_{{\rm gas},i}}{\Delta t}=A(t,R)+\sum_{j}F_{i,j}+\sum_{k}G_{i,k}.
\end{equation}

\begin{table}[ht]
\centering
\parbox[l]{95mm}{{\smallbf\ \ Table 1.}
{\small\ The parameter sources for the P\'{E}GASE software.\lstrut}}
\small
\begin{tabular}{lll}
\hline
Parameter  & Value  & Reference \\ [0.5ex]
\hline
IMF  &  --- & Kroupa (2002)\hstrut \\
SN\,II yields  & Model B  & Woosley \& Weaver (1995) \\
Fraction of close binary systems & 0.05 & P\'{E}GASE default value \\
\hline
\end{tabular}
\label{table:nonlin1}
\end{table}

\subsectionb{2.5}{Gas dynamics}

The galactic disk model, subdivided into isolated cells of $\sim$\,100~pc in
size, would be an unrealistic approach, as typical multi-supernovae powered
super-bubbles reach a similar size (e.g., Chu 2008).  The isolated cells
could also lead in some cases to the highly inhomogeneous metallicity of
interstellar matter in galaxy disks, which clearly contradicts
observations (Scalo \& Elmegreen 2004).

The gas dynamics, implemented in our model, is based on assumptions that
the equilibrium gas distribution is defined by an exponential disk
scale length and that every SF event in a cell is powerful enough to inflate
a super-bubble beyond the cell's boundaries, leaving in it only the
rarefied gas of high temperature.  Assuming the initial mass function
(IMF) (see Table 1) and the energy released by supernovae, $E_{\rm SN} =
10^{44}$~J, the super-bubble radius after 10~Myr is equal to
$\sim$\,140~pc (McCray \& Kafatos 1987) at the edge of the star forming
disk ($\sigma_{{\rm gas},i}=7\,M_{\odot}\ \textrm{pc}^{-2}$).  This
proves that SF events in the cells are powerful enough to drive
inter-cell gas dynamics.  As a consequence, the gas, flowing from the cell
$i$ to the cell $j$ is proportional to the mass of gas in the cell $i$
and the fraction of its perimeter, contacting the cell $j$:  $F_{i,j}
\sim m_{{\rm gas},i} \cdot \lambda_{i,j}$.

The cavity in the cell $i$, produced by a SF event and supported by a
SN\,II, the main driving mechanism of the super-bubble (Mac Low \&
McCray 1988), starts to vanish after about 40~Myr due to refilling with
gas through diffusion processes or SF events in the neighboring cells.
For our simulation we assume a diffusion timescale of 350~Myr for a cell
size region, following the numerical simulations by Recchi \& Hensler
(2005), who found the refill time to be ranging from 125 to 600~Myr.

\sectionb{3}{APPLICATION OF THE MODEL}

The described model is most appropriate for late-type spiral galaxies
for the following reasons.  Firstly, because the models based on the SSPSF
scenario produce patchy star forming region patterns, which are similar
to late-type flocculent spiral galaxies (Seiden \& Gerola 1982).
Secondly, we do not model the formation and evolution of the bulge,
therefore, the results are most adequate to galaxies with a negligible
bulge contribution.

According to those model properties, one of the best candidates for
model tests is the Local Group galaxy M\,33.  It is the nearest
late-type spiral having a moderate inclination (see Table 2 for the
adopted parameters).  As shown by Ferguson et al.  (2007) the lack of
substructures in the outer regions of this galaxy indicates its evolution in
relative isolation.  The small distance to the galaxy ensures plenty of
observational data and makes it the ideal target for chemical and
spectrophotometric evolution studies in terms of the proposed model.

\begin{table}[ht]
\centering
\parbox[l]{95mm}{{\smallbf\ \ Table 2.}{\small\ The parameters of M\,33\lstrut}}
\small
\begin{tabular}{lll}
\hline 
Parameter  & Value  & Reference \\ [0.5ex]
\hline
Morphological type  & Sc &  Paturel et al.\hstrut (2003)\footnotemark\\
Disk inclination  & $54^{\circ}$  & Paturel et al. (2003)\footnotemark[\value{footnote}]\\
Position angle of major axis  & $22.5^{\circ}$  & Paturel et al. (2003)\footnotemark[\value{footnote}]\\
Distance  & 840\,kpc  & Freedman et al. (1991)\\
Optical disk radius (\emph{B} = 25\,mag/arcsec$^2$)  & 7.3\,kpc  & Paturel et al. (2003)\footnotemark[\value{footnote}]\\
Baryonic mass  & $\leq$$10^{10}\,\textrm{M}_{\odot}$  & Corbelli (2003)\\
\hline
\end{tabular}\label{table:nonlin2}
\end{table}

\footnotetext[1]{http://leda.univ-lyon1.fr}

\subsectionb{3.1}{Model calibration}

We model M\,33 by simulating a disk composed of 99 rings.  A cell
width of 100~pc corresponds to a 10~kpc disk radius and fits well with the
optical disk radius of the galaxy.  The disk formation timescale was
chosen to correspond to a slow accretion scenario.  As recent studies have
shown, such a scenario is in better agreement with observations (Magrini
et al. 2007) and the up-to-date infall rate (Grossi et al. 2008).  An
accretion timescale, $\tau_{\rm acc}$, of 6~Gyr, typical to Sc-type
galaxies (Arimoto et al. 1992), and a disk scale-length of 2~kpc
(Freeman 1970), were applied.  The distribution of mass in the reservoir
was chosen to fit the baryonic mass radial distribution calculated from
the gas and star surface densities derived by Corbelli (2003).  The
galaxy rotation data are taken from Corbelli \& Salluci (2000).  Based
on the best estimate from our modeling of M\,33, we assume its age 
to be 12~Gyr.

\subsectionb{3.2}{Comparison with observations}

Our model predicts gas and star disk surface densities, which are in
good agreement with observations (Figures~2 and 3).  A good agreement 
with the gas surface density indicates that the adopted parametrization of
gas dynamics is adequate for this study.  However, the predicted 
relatively sharp cut-off in the stellar density profile could be an
artifact of our model, because there is no dispersion of stars between
the cells introduced, and stars remain {\it in situ} positions.
Ro{\v{s}}kar et al.  (2008) have shown that such an assumption is
incorrect, however, this does not change the galactic parameters
significantly.

\vskip2mm

%%%%%%%%%%%%%%%%%%%%%%%%%%  FIGURES 2, 3

\vbox{\centerline{\psfig{figure=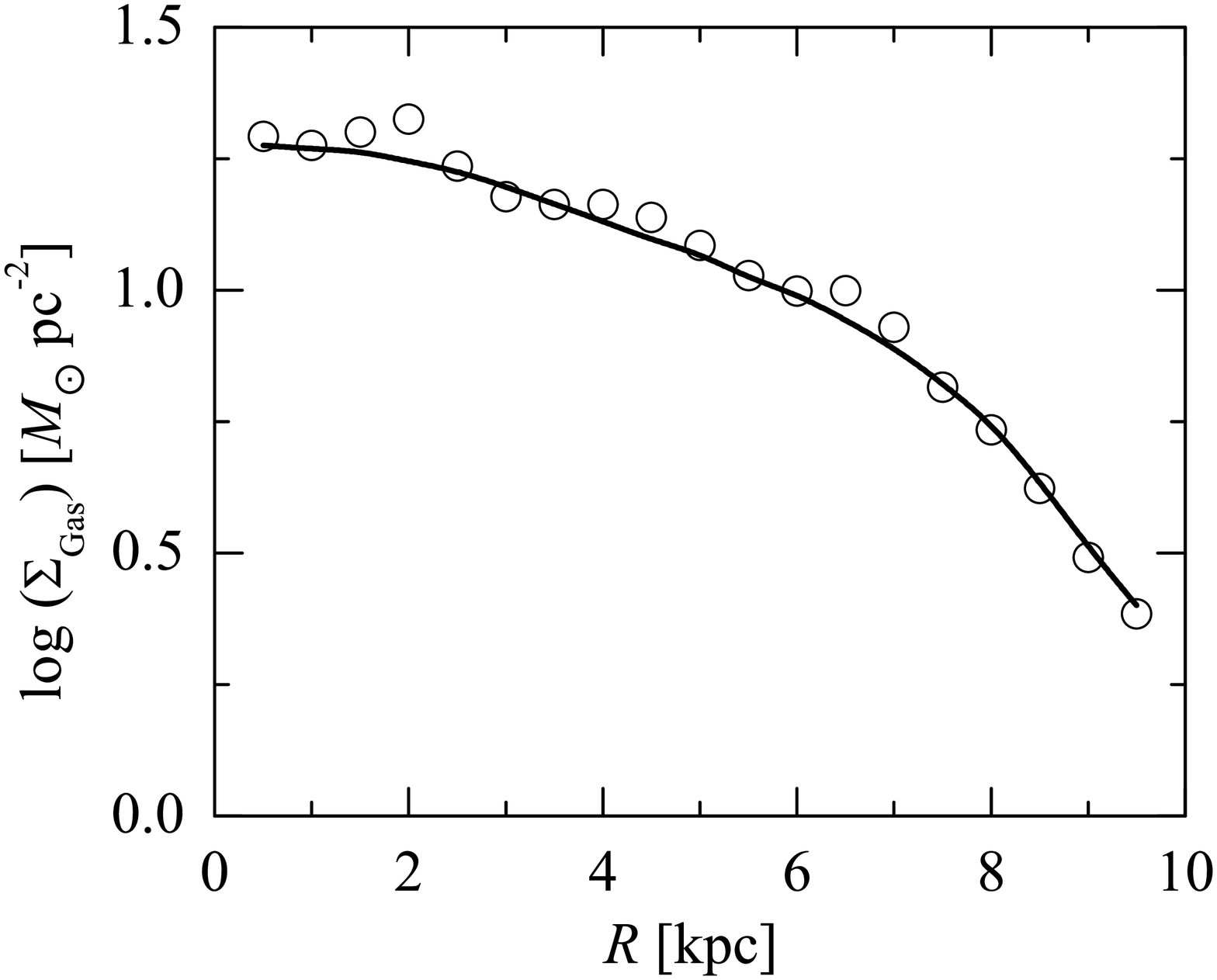,width=70truemm,angle=0,clip=}}
\vspace{0.5mm}
\captionb{2}{The profiles of gas surface density. The solid
line denotes the average of 30 models; the circles represent
observational data from Corbelli (2003).}}

\vskip6mm

\vbox{\centerline{\psfig{figure=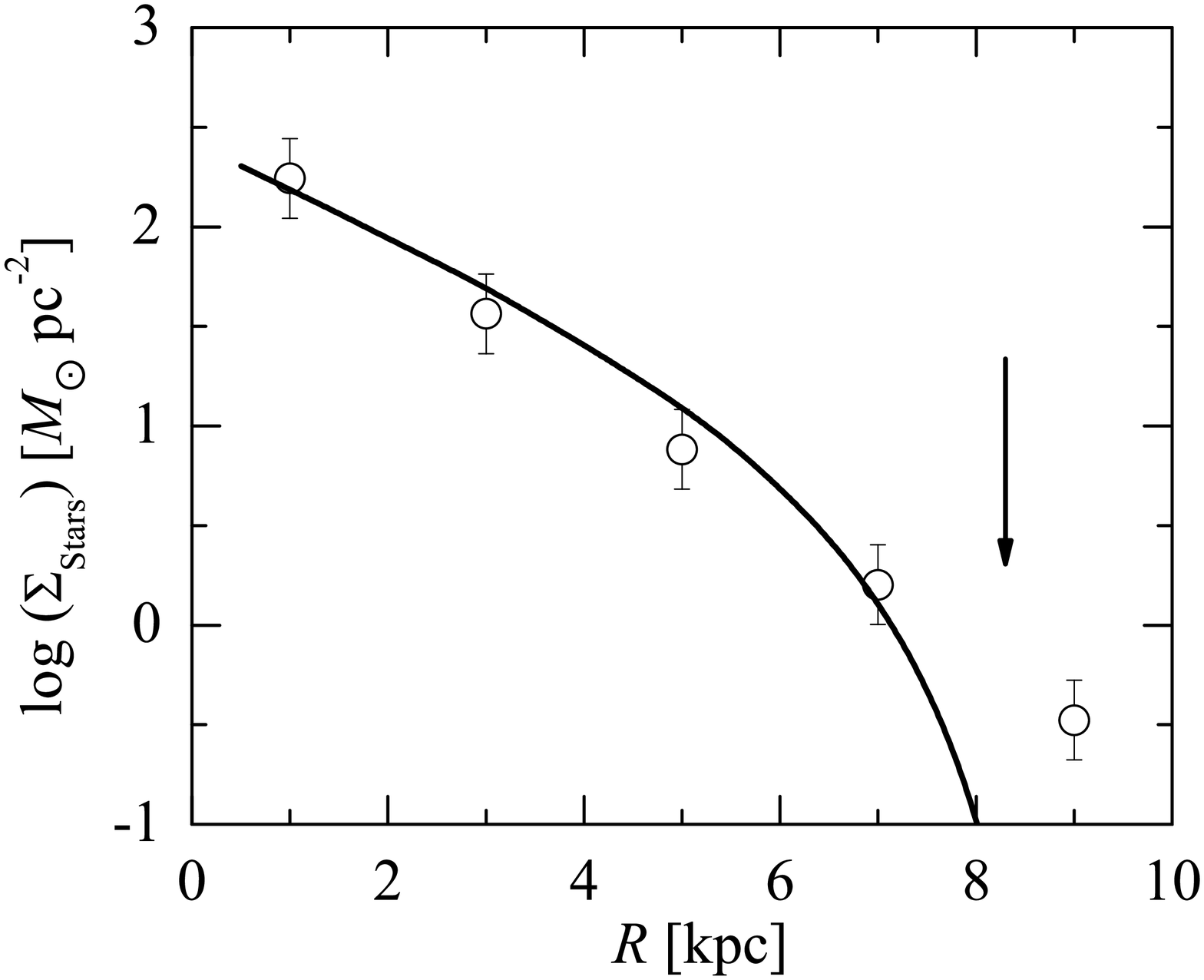,width=70truemm,angle=0,clip=}}
\vspace{0.5mm}
\captionb{3}{The surface density profiles of stars in the disk. The
solid line denotes the average
of 30 models; the circles represent observational data from Corbelli
(2003). The arrow indicates the location of a sharp break
in the stellar density profile (Ferguson et al. 2007).}}

\vskip3mm

The radial profile of star formation rate predicted by the model
(Figure~4) is also in good agreement with observations.

For the comparison of M\,33 with the model predictions we use oxygen
abundances in H\,II zones and metallicities of blue supergiants.  The
model oxygen abundances are derived assuming the scaled solar
metallicity by Asplund et al.  (2005).  Recently, Rosolowsky \& Simon
(2008) provided the largest homogeneous sample of H\,II zone abundances in
M\,33 (Figure~5).  The metallicities of blue supergiants (Figure~6) are
taken from Urbaneja et al.  (2005) and U et al.  (2009).  It is evident
that both samples display different behavior.  The supergiants show
a steeper metallicity gradient and systematically higher metallicity
values.  The determination of the radial metallicity gradient in
H\,II zones is uncertain due to a high intrinsic scatter of the data.

\vskip2mm

%%%%%%%%%%%%%%%%%%%%%%%%%%%%%%%  FIGURES 4, 5, 6

\vbox{\centerline{\psfig{figure=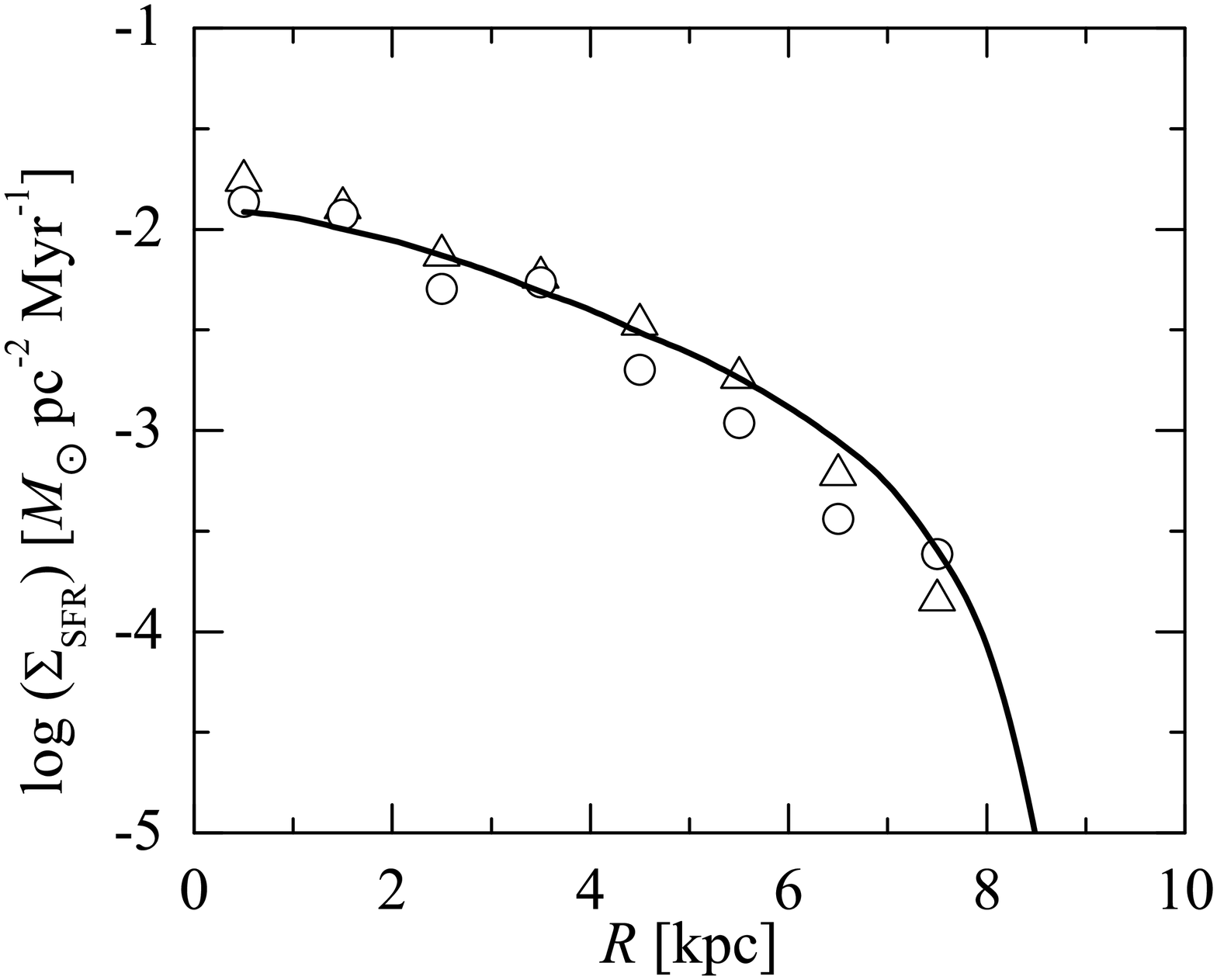,width=70truemm,angle=0,clip=}}
\vspace{0.5mm}
\captionb{4}{Star formation rate (SFR) density profiles.
The solid line denotes the average of 30 models; the triangles and
circles represent data derived from observations (Verley et al. 2009) using
different tracers (FUV and H$\alpha$ respectively).}}

\vskip2mm

\vbox{\centerline{\psfig{figure=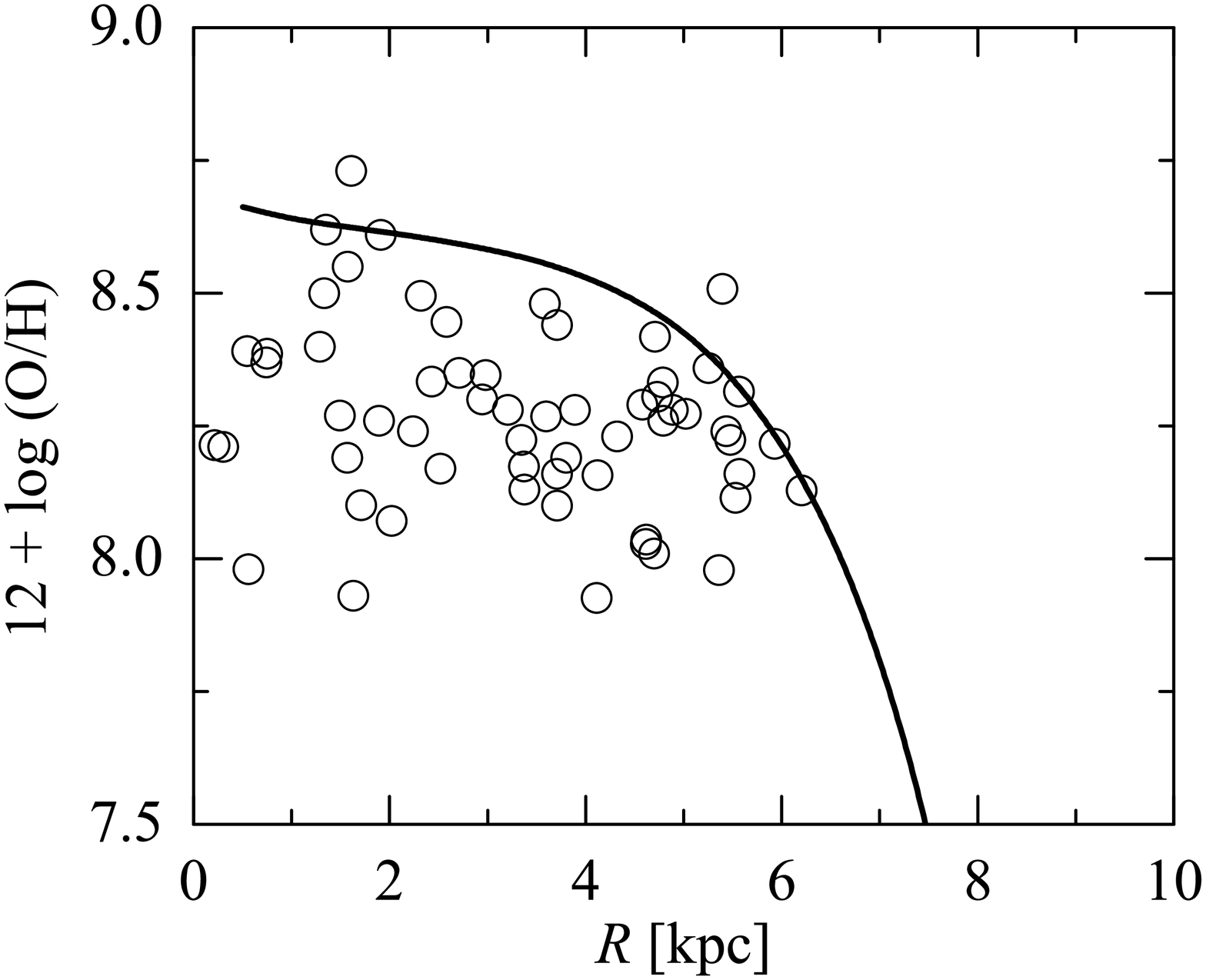,width=70truemm,angle=0,clip=}}
\vspace{0.5mm}
\captionb{5}{The oxygen abundance model profile of gas in M\,33.
The solid line denotes the average of 30 models; the circles represent
abundances of the H\,II zones (Rosolowsky \& Simon 2008).}}

\vbox{\centerline{\psfig{figure=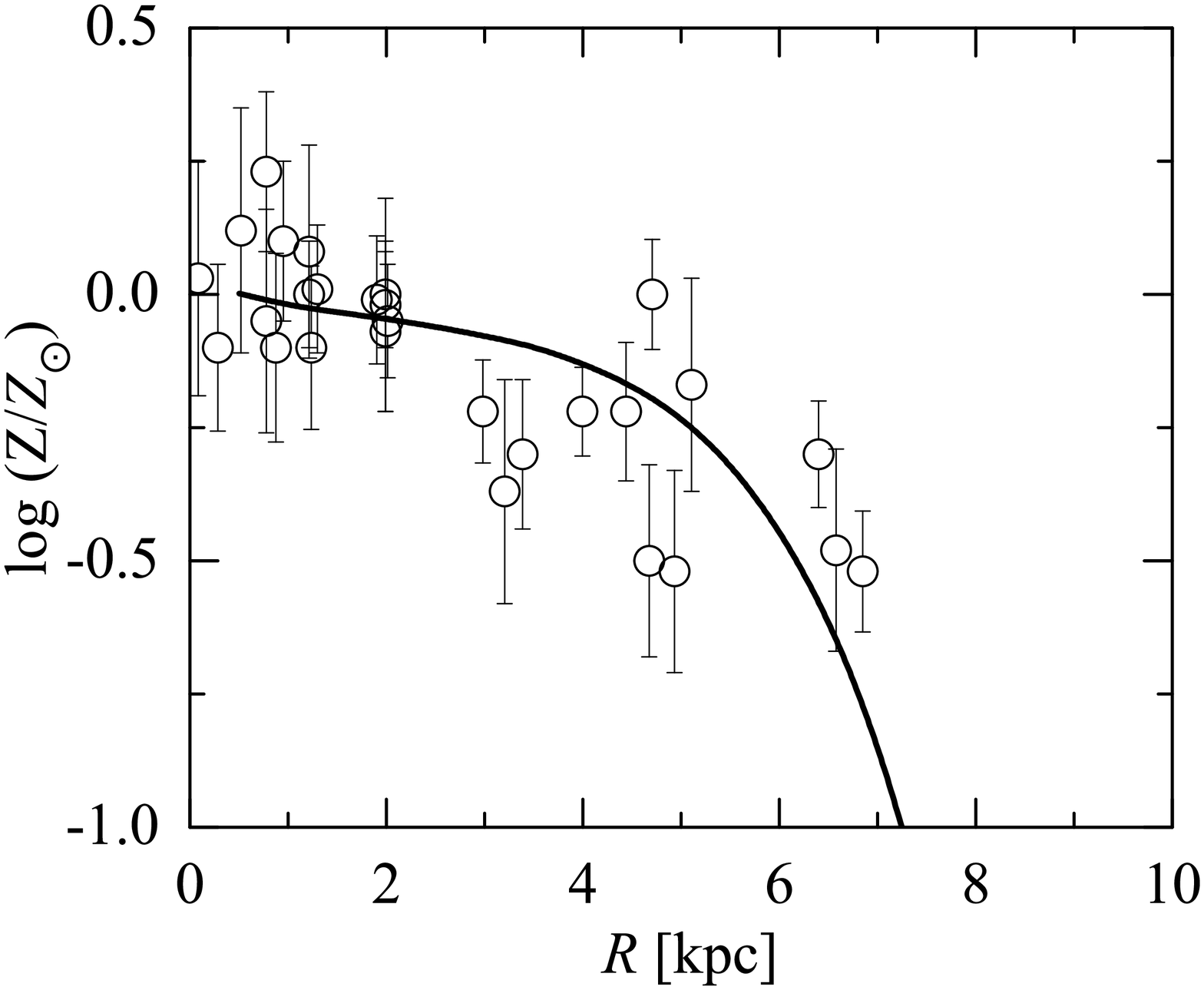,width=70truemm,angle=0,clip=}}
\vspace{0.5mm}
\captionb{6}{The metallicity profile of gas in M\,33.
The solid line denotes the average of 30 models; the circles with error
bars indicate the abundances in blue supergiants from Urbaneja et
al. (2005) and U et al. (2009).}}

\vskip3mm

A systematic H\,II zone ``under-abundance'' was discussed by Stasi\'nska
(2005) -- at the solar metallicity the measured abundance can be
underestimated by 0.2\,dex.  Therefore, bearing in mind systematic
differences in abundances between the H\,II zones and blue supergiants,
our model is in reasonable agreement with both datasets.  Judging from a
few observation data points (Urbaneja et al. 2005, U et al. 2009) at
large radii, our model produces too sharp cutoff in the outer disk,
which could be explained by the neglected star exchange between the
model's cells.  Spitoni et al.  (2008) have shown that during a SF event
there is a possibility to contaminate regions at distances of up to 1\,kpc
by a super-bubble blow-out which can considerably smooth the abundance
gradients in real disks.

\sectionb{4}{CONCLUSIONS}

We propose a new model of chemical and spectrophotometric evolution of
disk galaxies, based on the stochastic self-propagating star formation
scenario.  We have extended the disk galaxy model by Seiden \& Gerola
(1982) and supplemented it with the disk formation through the accretion
process, the parameterized gas dynamics inside the disk, and the
chemical and photometric evolution treatment based on a simple stellar
population approach.

The model of the late-type galaxy M\,33 is in good agreement with
the observed radial profiles of gas and star surface density, oxygen
abundance, metallicity and star formation rate surface density.

\References

\parskip=0.1pt

\refb Arimoto~N., Yoshii~Y., Takahara~F.\ 1992, A\&A, 253, 21
\refb Asplund~M., Grevesse~N., Sauval~A.~J.\ 2005, in \emph{Cosmic Abundances as Records of Stellar Evolution and Nucleosynthesis}, eds. T.~G.~Barnes III \& F.~N.~Bash, ASP Conf. Ser., 336, 25
\refb Bolatto~A.~D., Leroy A.~K., Rosolowsky~E., Walter~F., Blitz~L.\ 2008, ApJ, 686, 948
\refb Corbelli~E., Salucci~P. 2000, MNRAS, 311, 441
\refb Corbelli~E.\ 2003, MNRAS, 342, 199
\refb Chu~Y.-H. 2008, in \emph{Massive Stars as Cosmic Engines} (IAU Symp. 250), eds. F.~Bresolin, P.~A.~Crowther, \& J.~Puls, Cambridge University Press, Cambridge, p. 341
\refb Chiappini~C., Matteucci~F., Romano~D.\ 2001, ApJ, 554, 1044
\refb Ferguson~A., Irwin~M., Chapman~S., Ibata~R., Lewis~G., Tanvir~N.\ 2007, in \emph{Island Universes - Structure and Evolution of Disk Galaxies}, ed R.~S.~de~Jong, Dordrecht, p. 239
\refb Fioc~M., Rocca-Volmerange~B. 1997, A\&A, 326, 950
\refb Freedman~W.~L., Wilson~C.~D., Madore~B.~F.\ 1991, ApJ, 372, 455
\refb Freeman~K.~C.\ 1970, ApJ, 160, 811
\refb Gerola~H., Seiden~P.~E. 1978, ApJ, 223, 129
\refb Gerola~H., Seiden~P.~E., Schulman~L.~S. 1980, ApJ, 242, 517
\refb Grossi~M., Giovanardi~C., Corbelli~E., Giovanelli~R., Haynes~M.~P., Martin~A.~M., Saintonge~A., Dowell~J.~D. 2008, A\&A, 487, 161
\refb Hensler~G. 2009, in \emph{The Galaxy Disk in Cosmological Context} (IAU Symp. 254), eds. J.~Andersen, J.~Bland-Hawthorn, \& B.~Nordstr{\"o}m, Cambridge University Press, Cambridge, p. 269
\refb Jungwiert~B., Palou{\v s}~J.\ 1994, A\&A, 287, 55
\refb Kennicutt~R.\ 1998, ApJ, 498, 541
\refb K\"oppen~J., Theis~C., Hensler~G. 1995, A\&A, 296, 99
\refb Kroupa~P. 2002, Science, 295, 82
\refb Mac Low~M.-M., McCray~R.\ 1988, ApJ, 324, 776
\refb Magrini~L., Corbelli~E., Galli~D. 2007, A\&A, 470, 843
\refb McCray~R., Kafatos~M.\ 1987, ApJ, 317, 190
\refb Myers~P.~C., Dame~T.~M., Thaddeus~P., Cohen~R.~S., Silverberg~R.~F., Dwek~E., Hauser~M.~G.\ 1986, ApJ, 301, 398
\refb Palou{\v s}~J., Tenorio-Tagle~G., Franco~J.\ 1994, MNRAS, 270, 75
\refb Paturel~G., Petit~C., Prugniel~P., Theureau~G., Rousseau~J., Brouty~M., Dubois~P., Cambr{\'e}sy~L. 2003, A\&A, 412, 45
\refb Quillen~A.~C., Bland-Hawthorn~J. 2008, MNRAS, 386, 2227
\refb Recchi~S., Hensler~G.\ 2006, A\&A, 445, L39
\refb Rosolowsky~E., Simon~J.~D.\ 2008, ApJ, 675, 1213
\refb Ro{\v{s}}kar~R., Debattista~V.~P., Quinn~T.~R., Stinson~G.~S., Wadsley~J.\ 2008, ApJL, 684, L79
\refb Scalo~J., Elmegreen~B.~G. 2004, ARAA, 42, 275
\refb Schaye~J.\ 2004, ApJ, 609, 667
\refb Seiden~P.~E., Gerola~H. 1982, Fundamentals of Cosmic Physics, 7, 241
\refb Sleath~J.~P., Alexander~P. 1995, MNRAS, 275, 507
\refb Spitoni~E., Recchi~S., Matteucci~F.\ 2008, A\&A, 484, 743
\refb Stasi{\'{n}}ska~G.\ 2005, A\&A, 434, 507
\refb U~V., Urbaneja~M.~A., Kudritzki~R.-P., Jacobs~B.~A., Bresolin~F., Przybilla~N.\ 2009, ApJ, 704, 1120
\refb Urbaneja~M.~A., Herrero~A., Kudritzki~R.-P., Najarro~F., Smartt~S.~J., Puls~J., Lennon~D.~J., Corral~L.~J.\ 2005, ApJ, 635, 311
\refb Verley~S., Corbelli~E., Giovanardi~C., Hunt~L.~K.\ 2009, A\&A, 493, 453
\refb Weaver~R., McCray~R., Castor~J., Shapiro~P., Moore~R. 1977, ApJ, 218, 377
\refb Wolfire~M.~G., McKee~C.~F., Hollenbach~D., Tielens~A.~G.~G.~M. 2003, ApJ, 587, 278
\refb Woosley~S.~E., Weaver~T.~A. 1995, ApJS, 101, 181
\end{document}